\documentclass[11pt,reqno]{amsart}
\usepackage{graphicx}
\usepackage{dcolumn}
\usepackage{bm}
\usepackage{amssymb}
\usepackage{amsfonts}
\usepackage{amsmath}
\usepackage{amsthm}
\usepackage{yhmath}
\usepackage{slashed}
\usepackage{xcolor}
\usepackage{comment}
\usepackage{scalerel,stackengine}

\usepackage{epsfig}
\usepackage{mathtools}
\usepackage{cite}
\usepackage{mathalfa}

\newcounter{mnotecount}

\newcommand{\mnotex}[1]
{\protect{\stepcounter{mnotecount}}$^{\mbox{\footnotesize $\bullet$\themnotecount}}$ 
\marginpar{
\raggedright\tiny\em
$\!\!\!\!\!\!\,\bullet$\themnotecount: #1} }

\stackMath
\newcommand\reallywidehat[1]{%
\savestack{\tmpbox}{\stretchto{%
  \scaleto{%
    \scalerel*[\widthof{\ensuremath{#1}}]{\kern-.6pt\bigwedge\kern-.6pt}%
    {\rule[-\textheight/2]{1ex}{\textheight}}
  }{\textheight}%
}{0.5ex}}%
\stackon[1pt]{#1}{\tmpbox}%
}

\def\logconf{\log{\Omega}}
\def\boun{\partial\mathcal{S}}

\newtheorem{Prop}{Proposition}
\newtheorem{Def}{Definition}

\newtheorem{remark}{Remark}
\theoremstyle{remark} 

\newcommand{\smd}{\slashed{\mathcal{D}}}

\begin{document}

\title[Mass-quasilocal angular momentum inequality for initial data with MFTS]{New spinorial mass-quasilocal angular momentum inequality for initial data with marginally future trapped surface}

\author{Jaros\l aw Kopi\'nski}
\address{Center for Quantum Mathematics and Physics (QMAP),
  Department of Mathematics,
  University of California,
  Davis, CA95616, USA}
\email{jkop@math.ucdavis.edu}

\author{Alberto Soria}
\address{Escuela T\'ecnica Superior de Ingenieros de Telecomunicaci\'on (ETSIT),
Departamento de Matem\'atica Aplicada a las TIC,
Universidad Politécnica de Madrid,
Avenida Complutense 30, 28040, Madrid, Spain
}
\email{alberto.soria@upm.es}

\author{Juan A. Valiente Kroon}
\address{School of Mathematical Sciences, Queen Mary University of London, E1 4NS, London, UK}
\email{j.a.valiente-kroon@qmul.ac.uk}

\date{\today}

\maketitle

\begin{abstract}
We prove a new geometric inequality that relates the Arnowitt-Deser-Misner (ADM) mass of initial data to a quasilocal angular momentum of a marginally future trapped surface (MFTS) inner boundary. The inequality is expressed in terms of a 1-spinor, which satisfies an intrinsic first-order Dirac-type equation. Furthermore, we show that if the initial data is axisymmetric, then the divergence-free vector used to define the quasilocal angular momentum cannot be a Killing field of the generic boundary.

\end{abstract}

\section{Introduction}

Geometric inequalities arise naturally in General Relativity (GR) as relations involving quantities characterizing black holes, like mass, angular momentum, and horizon area. One of the most significant example of a geometric inequality is the positive (ADM) mass theorem. It was first proven by Schoen and Yau for the time-symmetric case in dimension three in \cite{SchoenYau79bis} and \cite{SchoenYau81} and later extended to dimension less than 8 in \cite{SchoenYau79}. On the other hand, Witten \cite{Wit81} proved the spinorial version of the theorem in dimension 3, and Bartnik realised that the proof could be easily extended to higher dimensions provided the manifold was spin \cite{Bartnik86}. Witten's spinorial version of the theorem  was extended to the case of initial data with trapped surfaces by Gibbons, Hawking, Horowitz, and Perry in \cite{GibHawHorPer83}. The spinorial approach was further adapted by Ludvigsen and Vickers in the context of the Bondi mass in \cite{LudVick82}. A refined version of the positivity of the ADM mass has been formulated by Penrose in the form of a lower bound on this quantity in terms of the horizon area of a black hole. If true, it would provide further evidence in favor of the weak cosmic censorship conjecture \cite{Pen73}. There exists a stronger version of the Penrose inequality involving the angular momentum of the initial data, namely
\begin{equation}
\label{angPenroseineq}
m \geq \left(\frac{\vert S \vert}{16\pi}+\frac{4\pi J^2}{\vert S \vert}\right)^{\frac{1}{2}},
\end{equation}
where $m$, $J$ and $\vert S \vert$ are the ADM mass, angular momentum and the outermost apparent horizon area respectively---see e.g. \cite{Mars2009, DainGabachC2018, Dain2014} for more details. This inequality is expected to hold only in axial symmetry. It admits a rigidity case, where equality exclusively occurs for the Kerr black hole. A quasilocal version of this relation states that
\begin{equation*}
m \geq\left(\frac{|S|}{16\pi}+\frac{4\pi J^2_{BH}}{\vert S \vert}\right)^\frac{1}{2},
\end{equation*}
where $J_{BH}$ is the quasilocal angular momentum of the horizon.

\medskip
Geometric inequalities for black holes remain a very active area of research with new interesting results being obtained. Among them is a bound on the ADM energy in terms of horizon area, angular momentum, and charge obtained by Jaracz and Khuri in \cite{JaraczKhuri2018}. In \cite{Anglada2018, Anglada2020} a different approach was considered by Anglada, who used the monotonic properties of the Geroch and Hawking energy along the inverse mean curvature flow in order to prove a Penrose-like inequality with angular momentum.  The first author and Tafel considered perturbations of Schwarzschild data and showed that \eqref{angPenroseineq} holds in this setting \cite{KopinskiTafel2020, KopinskiTafel20202}. Another refinement to the Penrose inequality with angular momentum has been proven by Alaee and Kunduri for 4-dimensional biaxially symmetric maximal initial data \cite{AlaeeKunduri2023}. Additionally, recent numerical results such as the ones obtained in \cite{KulczyckiMalec2021} give support to the validity of (\ref{angPenroseineq}) in the context of axial symmetry. The examples presented above are far from exhaustive, providing a glimpse into contemporary research in geometric inequalities in GR. We refer the reader to \cite{Mars2009,DainGabachC2018} for further references.

\medskip
In the present work a spinorial approach is used to obtain a geometric inequality involving the ADM mass of the initial data for the vacuum Einstein field equations and a quasilocal angular momentum (à la Szabados \cite{Sza06,Sza08}) of the MFTS inner boundary. It generalises the result presented  in \cite{KopinskiKroon2021} to the case of non-vanishing connection 1-form on the normal bundle of the boundary. The solvability of the boundary value problem for the so-called \emph{approximate twistor equation} is still an essential ingredient for deriving the main result. The existence of solution is used to obtain a basic mass inequality  \begin{equation}
    \label{masterineq}
    4 \pi m  \geq \sqrt{2} \oint_{\boun} \widehat{\phi}^A \gamma_A{}^B \smd_{BC} \phi^C dS,
\end{equation}
where $\smd_{AB}$ and $\gamma_{AB}$ are the 2-dimensional Sen connection and the complex metric on the boundary respectively (see below for details), while $\phi_A$ is a valence 1 spinor on $\boun$. The right-hand side of \eqref{masterineq} can be rewritten in terms of the inner null expansion $\theta^-$ of the boundary and the aforementioned angular momentum, provided that $\phi_A$ satisfies a certain first-order Dirac-type equation. Ultimately,
\begin{equation}
     \label{Sphericalmassinequalityi}
      4 \pi m \geq \sqrt{2} \oint \displaylimits_{\mathbb{S}^2} \rho '  |\widetilde{\phi}_0|^2\Omega d\mathbb{S}^2 + \frac{\kappa}{\sqrt{2}}  O\left[\widetilde{\phi}, U \right],  
\end{equation}
where $\rho'= - \tfrac{\theta^-}{2}$, $\widetilde \phi_A$ is a Dirac eigenspinor on $\mathbb{S}^2$, $\Omega$ is a conformal factor relating the metrics of $\boun$ and $\mathbb{S}^2$ and $O\left[\widetilde{\phi}, U \right]$ a quasilocal angular momentum depending on $\widetilde{\phi}_A$ and a rotation potential $U$ defined below. It should be noted that the integrals in \eqref{Sphericalmassinequalityi} are now taken with respect to the 2-sphere volume element. 

A natural symmetry associated with the angular momentum is the existence of axial Killing vector. Therefore, with such assumption we analyze a scenario where the quasilocal angular momentum is generated by such vector (on top of arising from a spinor $\phi_A$) and show that it is in fact impossible for a generic MFTS inner boundary $\boun$.

\medskip
The article is structured as follows: Section \ref{Secpreliminaries} provides a discussion of our main mathematical tools, in particular a new formalism for the $1+1+2$ decomposition of spinors. Section \ref{Secapproximatetwistoreq} is an adaptation of the result of \cite{KopinskiKroon2021} to the case of non-vanishing connection 1-form on the normal bundle of $\boun$. In Section \ref{Secmassineq} we present the main result of this work, a new mass-quasilocal angular momentum inequality for the initial data with a MFTS. In the last section we particularise our analysis to the axisymmetric setting and show that the divergence-free vector generating the quasilocal angular momentum cannot arise simultaneously from a first-order Dirac-type equation and be a Killing vector of the boundary.

In the following, 4-dimensional metrics are considered to have the signature $(+---)$. As a result, Riemannian 3- and 2-dimensional metrics will be negative definite. Whenever appropriate, we will expand spinorial expressions using either the Geroch-Held-Penrose (GHP) or Newman-Penrose (NP) formalism, following the conventions outlined in \cite{PenRin84}. Throughout this paper, we employ abstract index notation, with lowercase letters representing tensorial indices and uppercase letters representing spinorial indices. Bold font will be used to denote components in a basis.

\section{Preliminaries}
\label{Secpreliminaries}

\subsection{Basic setting}

 An initial data set $(\mathcal{S},h_{ab},K_{ab})$ for the vacuum Einstein field equations is said to be \emph{asymptotically Schwarzschildean} if the metric $h_{ab}$ and the second fundamental form $K_{ab}$ satisfy the decay conditions
\begin{subequations}
\begin{eqnarray}
&& h_{ab} = - \left(1+\frac{2m}{r}\right)\delta_{ab} + o_\infty(r^{-3/2}),\label{DecayAssumption1} \\
&& K_{ab} = o_\infty(r^{-5/2}), \label{DecayAssumption2}
\end{eqnarray}
\end{subequations}
 with $r^2 \equiv  (x^1)^2 + (x^2)^2 + (x^3)^2$,   $(x^\alpha)=(x^1,x^2,x^3)$ being asymptotically Cartesian coordinates and $m$ the ADM mass. In this work we assume that $\mathcal{S}$ has an inner boundary $\boun$ which is a topological 2-sphere and is equipped with a metric $\sigma_{ab}$. We consider a $1+1+2$ spinor formalism, first  proposed in \cite{Sza94a} by Szabados, and based on the use of $SL(2,\mathbb{C})$ spinors.  Maintaining the same philosophy as in \cite{KopinskiKroon2021}, the so-called  
$SU(2,\mathbb{C})$ spinors (or space spinors) introduced in \cite{Som80} will be essential to our purposes since they allow to work efficiently on spacelike hypersurfaces. For more information on the spinor formalism, we refer the reader to \cite{CFEBook,Ash91}.

Let $\tau^{AA'}$ be a spinorial counterpart of the orthogonal future vector $\tau^a$ to $\mathcal{S}$ such that $\tau_{AA'}\tau^{AA'}=2$. Likewise, we will denote a spinorial counterpart of the normal vector $\rho^a$ to $\boun$ on $\mathcal{S}$ as $\rho^{AA'}$ and assume that $\rho_{AA'}\rho^{AA'}=-2$. Let us choose $\rho_{AA'}$ so that it is pointing outwards, towards infinity. The spinors $\tau_{AA'}$ and $\rho_{AA'}$ are orthogonal ---i.e. $\tau_{AA'}\rho^{AA'}=0$. We consider dyads $\{o^A,\iota^A\}$ such that
\begin{eqnarray*}
   && \tau_{AA'}=o_A \overline{o}_{A'}+\iota_A \overline{\iota}_{A'}, \nonumber \\
   && \rho_{AA'}=o_A \overline{o}_{A'}-\iota_A \overline{\iota}_{A'}.
\end{eqnarray*}
The spinor $\tau_{AA'}$ is used to construct a space-spinor version of a given spinor. In particular,
\begin{equation*}
    \gamma_{AB}\equiv \tau_{(B}{}^{A'}\rho_{A)A'}
\end{equation*}
is the space-spinor version of $\rho_{AA'}$, also called the \emph{complex metric}. By construction, the complex metric can be understood as the space spinor version of the vector $\rho_{AA'}$, which is the spacelike normal to $\partial S$ on $S$ (with the normalization $\rho_{AA'}\rho^{AA'}=-2$). It satisfies $\gamma_{A}{}^{B}\gamma_{B}{}^C=\delta_A{}^C$ and can be expressed as
\begin{equation*}
    \gamma_{AB}=o_A \iota_B+o_B \iota_A
\end{equation*}
with the use of spin dyad.

The spinorial counterpart of the projection operator $\Pi_a{}^{b}$
onto the 2-dimensional surface $\boun$ can now be defined as
\begin{equation}
\begin{split}
    \Pi_{AA'}{}^{BB'}  & \equiv \delta_A{}^B \delta_{A'}{}^{B'} - \tfrac12 \tau_{AA'} \tau^{BB'} + \tfrac12 \rho_{AA'} \rho^{BB'} = \tfrac12 \left(\delta_A{}^B \delta_{A'}{}^{B'} - \gamma_{A}{}^B \overline{\gamma}_{A'}{}^{B'} \right).
\end{split}
\end{equation}

Similarly, the spinorial counterpart of the projector $T_{AA'BB'}$ onto $\mathcal{S}$ reads
\begin{equation*}
    T_{AA'}{}^{BB'} \equiv \tfrac12 \left(\delta_A{}^B \delta_{A'}{}^{B'} - \tfrac{1}{2}\tau_{AA'}\tau^{BB'} \right).
\end{equation*}
Let $\nabla_{AA'}$ be the spinorial counterpart of the spacetime covariant derivative $\nabla_a$.
The $T_{AA'}{}^{BB'}$ projector allows to define the $3$-dimensional Sen connection $\mathcal{D}_{AA'}$
associated to $\nabla_{AA'}$ as
\begin{equation*}
        \mathcal{D}_{AA'}\pi_C \equiv T_{AA'}{}^{BB'}\nabla_{BB'}\pi_C.
\end{equation*}
As mentioned above, the $SU(2,\mathbb{C})$ (i.e. space-spinor version) of $\mathcal{D}_{AA'}$ can be constructed by means of $\tau^{AA'}$ as
\begin{equation*}
    \mathcal{D}_{AB}=\tau_{(B}{}^{A'}\mathcal{D}_{A)A'}.
\end{equation*}

The space-spinor version $\nabla_{AB}$ of the 3-dimensional Levi-Civita connection on $\mathcal{S}$ can be recovered form $\mathcal{D}_{AB}$ via  
\begin{equation*}
\nabla_{AB}\pi_C=\mathcal{D}_{AB}\pi_C-\tfrac{1}{2}K_{ABC}{}^Q \pi_Q,
\end{equation*}
where $K_{ABCD} \equiv \tau_D{}^{C'}\mathcal{D}_{AB}\tau_{CC'}$ is the Weingarten spinor (note that symbol $D_{AB}$ was chosen in \cite{KopinskiKroon2021} to denote the space version of the 3-dimensional Levi-Civita connection. Here we prefer to keep the symbol $\nabla$ for Levi-Civita connections). The Weingarten spinor decomposes as
\begin{equation*}
    K_{ABCD}=\Omega_{ABCD}-\tfrac{1}{3}K\epsilon_{A(C}\epsilon_{D)B}, 
\end{equation*}
where $\Omega_{ABCD} \equiv K_{(ABCD)}$ is its fully symmetrized part, and $K\equiv K_{AB}{}^{AB}$ is the mean curvature of $\mathcal{S}$. The 3-dimensional Levi-Civita operator satisfies $\nabla_{AB}\epsilon_{CD}=0$.

Given a spinor $\pi_{A_1{}\dots{}A_K}$, its \textit{Hermitian conjugate} is defined as follows,
\begin{equation*}
    \widehat{\pi}_{A_1{}\dots{}A_K} \equiv\tau_{A_1}{}^{A_1'}\dots\tau_{A_k}{}^{A_k'} \, \overline{\pi}_{A_1'\dots A_k'}.
\end{equation*}
A spinor $\pi_{A_1{}\dots{}A_K}$ is said to be real if
\begin{equation*}
    \widehat{\pi}_{A_1 B_1{}\dots{}A_k B_k}{}^{C_1 D_1{}\dots{}C_m D_m}=(-1)^{(k+m)}\pi_{A_1 B_1{}\dots{}A_k B_k}{}^{C_1 D_1{}\dots{}C_m D_m}.
\end{equation*}

The space counterpart of the Levi-Civita connection $\nabla_{AB}$ is real in the sense that $ \widehat{\nabla_{AB} \pi_C} = - \nabla_{AB}\widehat{\pi}_C$, while
$$
\widehat{\mathcal{D}_{AB} \pi_C} = -\mathcal{D}_{AB} \widehat{\pi}_C +
   K_{ABC}{}^D \widehat{\pi}_D.
$$

\subsection{On the inner boundary.}

A 2-dimensional Sen connection $\slashed{\mathcal{D}}_{AA'}$ on $\boun$ arises as a $\Pi$-projection of $\nabla_{AA'}$, i.e.
\begin{equation}
    \label{2DSenonevalence}
    \slashed{\mathcal{D}}_{AA'} \equiv \Pi_{AA'}{}^{BB'}\nabla_{BB'}, 
\end{equation}
and its associated $SU(2,\mathbb{C})$ version is given by $ \slashed{\mathcal{D}}_{AB} \equiv \tau_{(B}{}^{A'}\slashed{\mathcal{D}}_{A)A'}$. It can be promoted to the 2-dimensional Levi-Civita connection $\slashed{\nabla}_{AA'}$ with the use of the transition spinor $Q_{AA'BC}$,
\begin{equation}
     \label{2DLCSentwovalence}
    \slashed{\nabla}_{AA'} v_{BB'} = \smd_{AA'} v_{BB'} - Q_{AA'B}{}^{C} v_{CB'} - \overline{Q}_{AA' B'}{}^{C'} v_{BC'},
\end{equation}
where
\begin{equation}
    Q_{AA'BC} \equiv - \tfrac12 \gamma_{C}{}^D \smd_{AA'} \gamma_{BD}.
\end{equation}
The $\slashed{\nabla}_{AA'}$ connection is torsion-free by definition, i.e.
$
   \left( \slashed{\nabla}_{AA'} \slashed{\nabla}_{BB'}  - \slashed{\nabla}_{BB'} \slashed{\nabla}_{AA'} \right) \phi = 0$,
and its curvature spinor $\slashed{r}_{CC'DD'AA'BB'}$ can be defined defined with the use of the following relation
\begin{equation}
     \left( \slashed{\nabla}_{AA'} \slashed{\nabla}_{BB'}  - \slashed{\nabla}_{BB'} \slashed{\nabla}_{AA'} \right) \pi^C= \slashed{r}^C{}_{QAA'BB'} \pi^Q =  \left(m_a \overline{m}_b - \overline{m}_a m_b \right) \left( \rho \rho' - \sigma \sigma' + \Psi_2 \right)\gamma^{CD} \pi_D,
\end{equation}
where $m^{AA'} \equiv o^A\overline{\iota}^{A'}$, $\overline{m}^{AA'} \equiv \iota^A \overline{o}^{A'}$, and 
$\Psi_2\equiv \Psi_{ABCD}o^A o^B \iota^C \iota^D$ is a component of the Weyl spinor $\Psi_{ABCD}$.

Another 2-dimensional connection ($\slashed{D}_{AB}$) can be obtained by considering a space-spinor counterpart of $\slashed{\nabla}_{AA'}$, i.e. 
\begin{equation}
\label{fakeLC}
    \slashed{D}_{AB} \equiv  \tau_{(B}{}^{A'} \slashed{\nabla}_{A)A'}.
\end{equation}
It is particularly useful in some calculations
and can be related to $\slashed{\mathcal{D}}_{AB}$ via 
\begin{equation}
\label{2DSennfakeLCrel}
\slashed{D}_{AB}\pi_C=\slashed{\mathcal{D}}_{AB}\pi_C-
Q_{AB}{}^Q{}_C\pi_Q,
\end{equation}
where the transition spinor is now given by
\begin{equation}
\label{trasitionQGHP}
Q_{AB}{}^C{}_D \equiv  -\tfrac{1}{2}\gamma_D{}^Q \slashed{\mathcal{D}}_{AB}\gamma_Q{}^C = \sigma' o_A o_B o_C o_D + \sigma \iota_A\iota_B\iota_C\iota_D - \rho o_A o_B \iota_C \iota_D - \rho' \iota_A \iota_B o_C o_D.
\end{equation}

A natural choice for the ingoing and outgoing null vectors $k^a$ and $l^a$ spanning the normal bundle to $\boun$ is given by
\begin{equation*}
    k^a=\tfrac{1}{2}(\tau^a-\rho^a), \qquad l^a=\tfrac{1}{2}(\tau^a+\rho^a).
\end{equation*}

The nature of a trapped surfaces is determined by the causal character and orientation of its mean curvature vector, or equivalently, by the signs of the associated inner and outer null expansions, 
\begin{equation*}
    \theta^{-}=\sigma^{ab}\nabla_a k_b, \qquad \theta^{+}=\sigma^{ab}\nabla_a l_b.
\end{equation*}
Making use of Proposition $4.14.2$ in \cite{PenRin84} we can express $\theta^-$ and $\theta^+$ in terms of a GHP spin coefficients, 
\begin{equation*}
    \theta^{-}=-2\rho' ,\qquad \theta^{+}=-2\rho.
\end{equation*}
We are now ready to define a marginally future trapped surface.

\begin{Def}
 The boundary $\boun$ is said to be a marginally future trapped surface (MFTS) if $\theta^{+}=0$ and $\theta^{-}\leq 0$ or if $\theta^{-}=0$ and $\theta^{+}\leq 0$, i.e. if $\rho=0$ and $\rho' \geq 0$ or if $\rho'=0$ and $\rho \geq 0$ on $\boun$.
 \end{Def}

\noindent
The 2-dimensional connection $\slashed{D}_{AB}$ annihilates $\epsilon_{AB}$ and $\gamma_{AB}$:
\begin{equation*}
   \slashed{D}_{AB}\epsilon_{CD}=0, \quad \slashed{D}_{AB}\gamma_{CD}=0.  
\end{equation*}
However, $\slashed{D}_{AB}$ is not a Levi-Civita connection on $\boun$ as it has a non-vanishing torsion,
\begin{equation} \label{2dcon}
\begin{split}
    \slashed{D}_{AB} \slashed{D}_{CD} \phi - \slashed{D}_{CD} \slashed{D}_{AB} \phi & =  \tfrac{1}{2} \left( A_{AB} \gamma_C{}^X \delta_D{}^Y - A_{CD} \gamma_A{}^X \delta_{B}{}^{Y} \right) \slashed{D}_{XY} \phi, 
\end{split}
\end{equation}
where $A_{AB}$ is defined as 
\begin{equation}
    \label{connectiononeform}
    A_{AB} \equiv  \tau^{CC'} \smd_{AB} \rho_{CC'} = 2 \left(\alpha + \overline{\beta} \right) o_A o_B - 2 \left(\overline{\alpha}+ \beta \right) \iota_A \iota_B. 
\end{equation}
Notice that $\slashed{D}_{AB}$ was considered in \cite{KopinskiKroon2021} to be the space version of the 2-dimensional Levi-Civita connection, being defined in the same way as in this work. This was possible since the boundary was torsion-free in \cite{KopinskiKroon2021} ($\alpha+\overline{\beta}=0$ on $\boun$). In the current work this restriction is dropped, and $A_{AB}\neq 0$.
This spinor is real, $\widehat{A}_{AB} = - A_{AB}$, and satisfies $\gamma^{AB} A_{AB}=0$. We can use it to recover a space-spinor version of the Levi-Civita connection $\slashed{\nabla}_{AB}$,
\begin{equation}
    \label{2DLCandDslashed}
    \slashed{\nabla}_{AB} \pi_C \equiv \slashed{D}_{AB} \pi_C - \tfrac{1}{4} A_{AB} \gamma_C{}^D \pi_D. 
\end{equation}
Indeed,
    \begin{equation} \label{slnabprop}
    \slashed{\nabla}_{AB} \epsilon_{CD} = \slashed{\nabla}_{AB} \gamma_{CD} = 0, \quad \gamma^{AB} \slashed{\nabla}_{AB} \pi_C = 0,
\end{equation}
and $\slashed{\nabla}_{AB}$ has vanishing torsion. Moreover, 
\begin{equation*}
  \left(    \slashed{\nabla}_{AC} \slashed{\nabla}_{B}{}^C -  \slashed{\nabla}_{B}{}^C \slashed{\nabla}_{AC} \right)\pi^B =  \tfrac{1}{2} \left( \Psi_2 + \rho \rho' - \sigma \sigma' + \mathrm{c.c.} \right) \pi_A,
    \end{equation*}
where c.c. denotes the complex conjugation of the expression in the brackets above. Ultimately, the 2-dimensional Sen and Levi-Civita connections on the boundary are related in the following way, 
\begin{equation}
\label{2dsenlc}
 \slashed{\mathcal{D}}_{AB}\phi_C=\slashed{\nabla}_{AB}\phi_C +\tfrac{1}{4}A_{AB}\gamma_C{}^D\phi_D+Q_{AB}{}^L{}_C\phi_L. 
\end{equation}
The connection $\slashed{\nabla}_{AB}$ is real, i.e.
\begin{equation*}
   \widehat{\slashed{\nabla}_{AB} \pi_C} = - \slashed{\nabla}_{AB}\widehat{\pi}_C, 
\end{equation*}
while,
$$ \widehat{\slashed{\mathcal{D}}_{AB} \pi_C}= -\slashed{\mathcal{D}}_{AB}\widehat{\pi}_C
   + \left(Q_{ABC}{}^D + \widehat{Q}_{ABC}{}^D \right) \widehat{\pi}_D+\tfrac{1}{2}\gamma_{C}^{\phantom{C}D}\widehat{\pi}_D A_{AB}.
$$
In the following we will also require an expression for the Hermitian conjugate of the connection $\slashed{D}_{AB}$. A direct computation yields
\begin{equation}
    \label{hermitfakeLC}
    \widehat{\slashed{D}_{AB}\pi_C}=-\slashed{D}_{AB}\widehat{\pi}_C+\tfrac{1}{2}\gamma_C{}^D\widehat{\pi}_D A_{AB}.
\end{equation}

In \cite{Sza06}, Szabados proposed the following definition of quasilocal angular momentum associated with $\boun$,
\begin{equation} \label{szab}
    O[N] \equiv  - \frac{1}{2\kappa} \oint_{\boun} N^c A_c dS,
\end{equation}
where $N^a$ is a divergence-free vector on $\boun$, $A_c=\rho_a \Pi^f{}_c\nabla_f\tau^a$ is the connection 1-form on the normal bundle of $\boun$ and $\kappa=8\pi G$ the gravitational coupling constant. After inspecting the definition \eqref{connectiononeform} of a spinor $A_{AB}$ we immediately see that it is in fact a space-spinor counterpart of $A_b$ in the expression above.

\medskip
In the sequel we will also make use of a Hodge decomposition on $\boun$. Specifically, given any 1-form $V_a$ on $\boun$ there exist two functions $f$ and $f'$ such that
\begin{equation*}
    V_a = \epsilon_a{}^b \slashed{\nabla}_b f +  \slashed{\nabla}_a f', 
\end{equation*}
where $\epsilon_{ab}$ is the $2$-dimensional Riemannian volume form of the boundary $\boun$.

\subsection{Conformal rescaling of the 2-dimensional Dirac operator} \label{subconfres}

An action of a 2-dimensional (Levi-Civita) \emph{Dirac operator} on a spinor $\pi_A$ is given by $\slashed{\nabla}_A{}^B \pi_B$. The purpose of this subsection is to explore its properties under conformal rescalings of the inner boundary metric $\sigma_{ab}$. Indeed, according to the uniformization theorem for compact Riemannian surfaces (\emph{cf.} \cite[Theorem 4.4.1]{Jost2006}), $\sigma_{ab}$ is conformal to the spherical metric, i.e.
\begin{equation*}
    \sigma_{ab}=\Omega^2 \widetilde{\sigma}_{ab}, 
\end{equation*}
where $\Omega: \mathbb{S}^2 \rightarrow \mathbb{R}$ is a non-negative smooth 
function and $\widetilde{\sigma}_{ab}$ a round 2-sphere metric. Given any holonomic basis $\{{\partial}_{{\bm A}{\bm B}}\}$ on $\boun$, the 
covariant derivative of $\pi_A$ can be expressed as
$\slashed{\nabla}_{AB}\pi_C=\partial_{{\bm A}{\bm B}}(\pi_{\bm C})-\Gamma_{{\bm A}{\bm B}}{}^{\bm Q}{}_{\bm C} \pi_{\bm Q}$,
where $\Gamma_{{\bm A}{\bm B}}{}^{\bm Q}{}_{\bm C} \equiv \frac{1}{2}\Gamma_{{\bm A}{\bm B}}{}^{{\bm Q}{\bm L}}{}_{{\bm C}{\bm L}}$ are the spin coefficients associated to the Christoffel symbols of $\sigma_{ab}$. Since $\sigma_{ABCD}=\Omega^2\widetilde{\sigma}_{ABCD}$, one arrives at
\begin{equation}
  \slashed{\nabla}_{AB}\pi_C=\widetilde{\slashed{\nabla}}_{AB}\pi_C -\tfrac{1}{2}\left(\partial_{AB}(\logconf)\widetilde{\sigma}^{EF}{}_{CF}+\partial_{CF}(\logconf)\widetilde{\sigma}^{EF}{}_{AB}-\partial^{EF}(\logconf)\widetilde{\sigma}_{ABCF}\right)\pi_E,
    \label{covderrelation}
\end{equation}
where $\widetilde{\slashed{\nabla}}_{AB}$ is a space-spinor counterpart of the Levi-Civita connection on $\mathbb{S}^2$. Contracting the second and third indices in the above with $\widetilde{\epsilon}^{AB}=\Omega^{-1}\epsilon^{AB}$ gives
    \begin{equation*}
        \slashed{\nabla}_A{}^B \pi_B=\Omega^{-1}\Big{[}\widetilde{\slashed{\nabla}}_{A}{}^B\pi_B -\frac{1}{2}( \partial_{A}{}^B(\logconf)\widetilde{\sigma}^{EF}{}_{BF} +\partial_{BF}(\logconf)\widetilde{\sigma}^{EF}{}_{A}{}^B-\partial^{EF}(\logconf)\widetilde{\sigma}_{A}{}^B{}_{BF})\pi_E\Big{]}, 
    \end{equation*}
with $\partial_{AB}$ defined as
\begin{equation}
\label{partialmatrix}
\partial_{AB}\equiv\frac{1}{\sqrt{2}}\begin{pmatrix}
-\partial_{x^1}-i\partial_{x^2} & \partial_{x^3} \\
\partial_{x^3} & \partial_{x^1}-i\partial_{x^2}  
\end{pmatrix},
\end{equation}
where the same the same convention for the Infeld-Van der Waerden symbols as in \cite{BaeVal11a} has been assumed. However, because $\widetilde{\sigma}_{ABCD}=\frac{1}{2}(\widetilde{\epsilon}_{AC}\widetilde{\epsilon}_{BD}+\widetilde{\gamma}_{AC}\widetilde{\gamma}_{BD})$ and
    $\widetilde{\gamma}^{AB}e_{AB}(\log\Omega)=0$ ($\widetilde{\gamma}_{AB}$ is orthogonal to $\boun$) the equation above reduces to
\begin{equation}
     \label{relconfDiracoperators}
    \slashed{\nabla}_A{}^B \pi_B=\Omega^{-1}\widetilde{\slashed{\nabla}}_{A}{}^B\pi_B.
\end{equation}
This can be used to establish the 
following fact: if $\widetilde{\pi}_A$ is a Dirac eigenspinor on $\mathbb{S}^2$ with an eigenvalue $\lambda\in\mathbb{R}$, i.e.
\begin{equation*}
    \widetilde{\slashed{\nabla}}_A{}^B \widetilde{\pi}_B = i \lambda \widetilde{\pi}_A,
\end{equation*}
then
\begin{equation}
    \label{confDirac}
    \slashed{\nabla}_A{}^B \pi_B = i \tfrac{\lambda}{\Omega} \pi_A,
\end{equation}

where the spin basis transforms in a following way, $\widetilde{o}_A=\tfrac{1}{\sqrt{\Omega}}o_A, \ \widetilde{\iota}_A=\tfrac{1}{\sqrt{\Omega}}\iota_A$ and $\widetilde{\pi}_0=\sqrt{\Omega}\pi_0, \  \widetilde{\pi}_1=\sqrt{\Omega}\pi_1 $.

\section{Approximate twistor equation}
\label{Secapproximatetwistoreq}
\subsection{Setup}

Let $\mathfrak{S}_1$, $\mathfrak{S}_3$ be the spaces of symmetric valence 1 and 3 spinors over the hypersurface $\mathcal{S}$. The (overdetermined) spatial twistor operator can be defined as follows,
\[
\mathbf{T}: \mathfrak{S}_1\rightarrow \mathfrak{S}_3, \qquad
\mathbf{T}(\kappa)_{ABC} \equiv \mathcal{D}_{(AB}\kappa_{C)},
\]  
and is a space-spinor counterpart of the twistor operator $\nabla_{A'(A}\kappa_{B)}$ (see \cite{BaeVal11a} for more details). 
The formal adjoint of $\mathbf{T}$ is given by 
\[
 \mathbf{T}^*: \mathfrak{S}_3\rightarrow \mathfrak{S}_1, 
\qquad \mathbf{T}^*(\zeta)_A
 \equiv  \mathcal{D}^{BC}\zeta_{ABC} - \Omega_A{}^{BCD}\zeta_{BCD}, 
\]
and allows to define the \emph{approximate twistor operator} $\mathbf{L} \equiv 
\mathbf{T}^*\circ\mathbf{T}:\mathfrak{S}_1\rightarrow\mathfrak{S}_1$,
\begin{equation}
\mathbf{L}(\kappa)_A \equiv
\mathcal{D}^{BC}\mathcal{D}_{(AB}\kappa_{C)}
-\Omega_A{}^{BCD}\mathcal{D}_{BC}\kappa_D,
\label{ApproximateTwistorEquation}
\end{equation}
which is formally self-adjoint ---i.e. $\mathbf{L}^*=\mathbf{L
}$. 

Let $\kappa_A$ be a solution of the approximate twistor equation $\mathbf{L}(\kappa)_A=0$. The spinors
\begin{equation*}
    \xi_A \equiv  \tfrac{2}{3}\mathcal{D}_A{}^Q\kappa_Q, \quad \xi_{ABC} \equiv  \mathcal{D}_{(AB}\kappa_{C)}
\end{equation*}
encode independent components of $\mathcal{D}_{AB}\kappa_C$. Moreover, one has that 
\begin{equation*}
    \mathbf{L} ( \widehat{\xi} )_A=0.
\end{equation*}

 Given the set of asymptotically Cartesian coordinates $(x^\alpha)$ on $\mathcal{S}$, the position spinor can be defined as follows,
$$
x_{\mathbf{A}\mathbf{B}} \equiv  \frac{1}{\sqrt{2}}
\left( 
\begin{array}{cc}
x^1+\mbox{i}x^2 & -x^3 \\
-x^3 & -x^1+\mbox{i} x^2
\end{array}
\right).
$$
We will consider a solution of the approximate twistor equation on \emph{asymptotically Schwarzschildean} initial data for the vacuum Einstein field equations with an asymptotic behaviour of the form 
\begin{equation}
\kappa_\mathbf{A} =\bigg( 1+\frac{m}{r} \bigg) x_{\mathbf{A}\mathbf{B}} o^{\mathbf{B}} +
o_\infty(r^{-1/2}).
\label{AsymptoticBehaviourSpinor}
\end{equation}
A direct computation shows that
\begin{subequations}
\begin{eqnarray}
&& \xi_{\mathbf{A}} =\bigg (1 -\frac{m}{r}\bigg) o_{\mathbf{A}} + o_{\infty}(r^{-3/2}), \label{DevelopXi1}\\
&& \xi_{\mathbf{ABC}} = -\frac{3m}{2r^3} x_{(\mathbf{AB}}
   o_{\mathbf{C})} + o_\infty(r^{-5/2}). \label{DevelopXi2}
\end{eqnarray}
\end{subequations}
 
As a consequence of the above asymptotic expansion of $\kappa_A$, one arrives at the following inequality relating the ADM mass of $\mathcal{S}$ and an integral of concomitants of the spinor $\kappa_A$, provided that the inner boundary $\boun$ is a MFTS \cite{KopinskiKroon2021},
\begin{equation}
4\pi m \geq \oint_{\partial \mathcal{S} } n_{AB} \zeta_C \widehat{\mathcal{D}^{(AB} \zeta^{C)}} \mbox{d} S,  
\label{BasicInequality}
\end{equation}
where $n_{AB}$ is the outer directed  (i.e. towards $r=\infty$) unit normal on $\boun$ as 
a surface of $\mathcal{S}$ and $\zeta_A\equiv \widehat{\xi}_A$.
 Since the space version of $\rho_{AA'}$ is the complex metric $\gamma_{AB}\equiv \tau_{(B}{}^{A'}\rho_{A)A'}$, relation $n_{AB}=\gamma_{AB}/\sqrt{2}$ holds. In the sequel we will use a boundary condition for $\kappa_A$ to refine the inequality \eqref{BasicInequality}.

\subsection{A boundary value problem for the approximate twistor equation}
Let
\begin{equation}
\label{boundarycondition}
\mathcal{D}_A{}^Q\kappa_Q=-\frac{3}{2}\widehat{\phi}_A \quad \mathrm{on} \quad \partial \mathcal{S},
\end{equation}
where $\phi_A$ is a smooth spinorial field. The approximate twistor equation together with \eqref{boundarycondition} satisfy the \emph{Lopatinskij-Shapiro} compatibility conditions (see eg. \cite{Dai06,WloRowLaw95}). This  implies that the associated boundary value problem is elliptic. Moreover, the decay conditions (\ref{DecayAssumption1}) and (\ref{DecayAssumption2}) for the first and second fundamental forms of the initial data 
make the approximate twistor operator $\mathbf{L}$ asymptotically homogeneous. In the sequel we will make use of an operator $\mathbf{B}$, defined in the in the following way,
\[
\mathbf{B}: \mathfrak{S}_1\rightarrow \mathfrak{S}_1, \qquad
\mathbf{B}(\kappa)_A \equiv  -\sqrt{2} \gamma_A{}^P \xi_P =
-\frac{2\sqrt{2}}{3}\gamma_A{}^P \mathcal{D}^Q{}_P\kappa_Q.
\] 
The equation (\ref{boundarycondition}) now becomes
\begin{equation*}
    \mathbf{B}(\kappa)_A{}|_{\boun}=\sqrt{2}\gamma_A{}^P\widehat{\phi}_P,
\end{equation*}
and the associated boundary value problem is
\begin{equation} \label{bvp}
    \mathbf{L}(\kappa)_A=0, \quad \mathbf{B}(\kappa)_A{}|_{\boun}=\sqrt{2}\gamma_{A}{}^P\widehat{\phi}_P.
\end{equation}

To discuss the solvability of \eqref{bvp} one has to look at the adjoint operators $\mathbf{L}^{*}$ and $ \mathbf{B}^*$. A similar computation as in \cite{KopinskiKroon2021} (in this case the extrinsic geometry of the boundary is non-trivial) leads to the following:

\begin{Prop}
If $\partial\mathcal{S}$ is a MFTS on the \emph{asymptotically Schwarzschildean} initial data set $(\mathcal{S},h_{ab},K_{ab})$ for the vacuum Einstein field equations, then the boundary value problem 
 \[
\mathbf{L}(\kappa)_A =0, \qquad \mathbf{B}(\kappa)_A|_{\partial\mathcal{S}}= \sqrt{2} \gamma_A{}^P \hat{\phi}_P,
\]
with a smooth spinorial field $\phi_A$ over $\partial \mathcal{S}$
admits a unique solution of the form
 \begin{equation}
\kappa_A =\mathring{\kappa}_A + \theta_A, \qquad \theta_A \in
H^2_{-1/2},
\label{Ansatz:SolutionApproximateTwistor}
\end{equation}
with $\mathring{\kappa}_A$ given by the leading term in
\eqref{AsymptoticBehaviourSpinor} and where $H^s_\beta$ with $s\in\mathbb{Z}^+$ and $\beta\in \mathbb{R}$ denotes the weighted $L^2$ Sobolev spaces.
\end{Prop}
\subsection{Inequality with the connection 1-form on the normal bundle of the boundary.}
The boundary condition \eqref{boundarycondition} allows to simplify the inequality \eqref{BasicInequality} to the following form
\begin{equation}
    \label{Essentialmassineq}
    4 \pi m \geq \sqrt{2} \oint_{\boun} \widehat{\phi}^A \gamma_A{}^B \smd_{BC} \phi^C dS,
\end{equation}
where $\phi_A$ is a free data in the boundary value problem \eqref{bvp}. All quantities in the integral are now intrinsic to the boundary. A relation \eqref{2dsenlc} between the 2-dimensional Sen and Levi-Civita connections implies that
\begin{equation*}
  \widehat{\phi}^A \gamma_A{}^B \smd_{BC} \phi^C=\widehat{\phi}^A \gamma_A{}^B (\slashed{\nabla}_{BC} \phi^C)+\tfrac{1}{4}\widehat{\phi}^A\gamma_A{}^B A_{BC}\, \gamma^{CL}\phi_L +\widehat{\phi}^A \gamma_A{}^B Q_{BC}{}^{LC}\phi_L.  
\end{equation*}
 Using \eqref{trasitionQGHP}, the GHP expression for $Q_{ABCD}$, it is easy to see that $Q_{AB}{}^{CB}=\rho'\iota_A o^C$ on a MFTS. This can be combined with the fact that $\gamma_A{}^B A_B{}^C\gamma_C{}^D=-A_A{}^D$ 
 to yield
\begin{equation}
\begin{split}
     \label{GHPmassinequality}
     4 \pi m \geq &  - \sqrt{2} \oint_{\boun} \widehat{\phi}^A \gamma_A{}^B (\slashed{\nabla}_B{}^C \phi_C)\,  dS + \sqrt{2} \oint_{\boun} \rho' |\phi_0|^2   dS - \frac{1}{2 \sqrt{2}} \oint_{\boun} A_{AB} \widehat{\phi}^A \phi^B  d S. 
\end{split}
\end{equation}
It should be noted that since $A_{AB}$ is intrinsic to $\boun$ the last term can be expressed as
\begin{equation} \label{angmom}
    - \frac{1}{2 \sqrt{2}} \oint_{\boun} \widehat{\phi}^A \phi^B \sigma_{AB}{}^{CD} A_{CD} d S,
\end{equation}
where $\sigma_{ABCD}$
$=\frac{1}{2} ( \epsilon_{AC} \epsilon_{BD} $
$ + \gamma_{AC} \gamma_{BD} )$ is the spinorial counterpart of the 2-dimensional metric.

\section{Mass-quasilocal angular momentum inequality}
\label{Secmassineq}

In this section we present the main result of this article -- the mass-quasilocal angular momentum inequality for the \emph{asymptotically Schwarzschildean} initial data $(\mathcal{S},h_{ab},K_{ab})$ for the vacuum Einstein field equations. It is based on a simplification of \eqref{GHPmassinequality} under suitable choice of the boundary spinor $\phi_A$. A natural condition for $\phi_A$ arises after inspecting the first term on the right-hand side of \eqref{GHPmassinequality} -- its 2-dimensional Dirac derivative should be controlled. Indeed, we will proceed with a following choice,
\begin{eqnarray}
\label{phigauge}
    \slashed{\nabla}_A{}^B \phi_B = i \frac{\lambda}{\Omega} \phi_A,
\end{eqnarray}
where $\Omega$ is a conformal factor relating a metric $\sigma_{ab}$ on $\boun$ with that on a round sphere $\mathbb{S}^2$. It can be showed that with a suitable choice of the conformal rescaling of the spin basis the equation \eqref{phigauge} corresponds to a Dirac eigenproblem on $\mathbb{S}^2$ (see Subsection \ref{subconfres} for more details).

The inequality (\ref{GHPmassinequality}) can now be simplified with the use of \eqref{phigauge} to the following form, 
\begin{equation}
     \label{GHPmassinequalityvanishterm}
     4 \pi m \geq  \sqrt{2} \oint_{\boun}  \rho' |\phi_0|^2  dS- \frac{1}{2 \sqrt{2}} \oint_{\boun} \widehat{\phi}^A \phi^B \sigma_{AB}{}^{CD} A_{CD} \, dS,
\end{equation}
where the reality of the ADM mass $m$ has been used to eliminate a (purely imaginary) term with the eigenvalue $\lambda$, i.e.
\begin{equation}
     \label{confcompatibilitycond}
\lambda \oint_{\boun}\left(|\phi_0|^2-|\phi_1|^2 \right)\Omega^{-1}dS=0.
\end{equation}

To make a connection between the second term on the right-hand side of \eqref{GHPmassinequalityvanishterm} and the quasilocal angular momentum \eqref{szab} we will introduce a spinor $N^{AB}$, defined as follows,
\begin{equation}
    \label{spindivfreevector}
    N^{AB} \equiv  \sigma^{ABCD} \phi_C \widehat{\phi}_D = \phi_0 \overline{\phi}_1 \iota^A \iota^B - \phi_1 \overline{\phi}_0 o^A o^B.
\end{equation}
One can verify that $N^{AB}$ is real, i.e. $\widehat{N}^{AB} = - N^{AB}$, so it corresponds to a real 3-vector. Moreover, $\gamma^{AB} N_{AB} =0 $ and
\begin{eqnarray}
   \slashed{\nabla}_a N^a &=& \slashed{\nabla}_{AB} \left( \sigma^{ABCD} \phi_C \widehat{\phi}_D \right) = \slashed{\nabla}_{AB} \left( \phi^A \widehat{\phi}^B \right) \nonumber \\
   &=& -(\slashed{\nabla}_B{}^A\phi_A)\widehat{\phi}^B+\phi^A (\widehat{\slashed{\nabla}_A{}^B\phi_B})=0, 
\end{eqnarray}
where (\ref{phigauge}) has been used in the last equality. Hence, $N^a$ is intrinsic to the boundary $\boun$ and $\slashed{\nabla}$-divergence-free, so we can identify it with $N^a$ generating the quasilocal angular momentum \eqref{szab}. With this choice the inequality \eqref{GHPmassinequalityvanishterm} yields
\begin{equation}
\label{masang1}
     4 \pi m \geq  \sqrt{2} \oint_{\boun}  \rho' |\phi_0|^2  dS+  \frac{\kappa}{\sqrt{2}} O\left[ \sigma^{ABCD} \phi_C \widehat{\phi}_D \right].
\end{equation}
In the remainder of this section we will simplify \eqref{masang1} and express it in terms of integrals over a round sphere $\mathbb{S}^2$ and the eigenspinor of the $\mathbb{S}^2$-Dirac operator.

The Hodge decomposition can be applied to the connection 1-form on the normal bundle of the boundary $A_b$ to yield $A_b = \epsilon_b{}^c \slashed{\nabla}_c U +  \slashed{\nabla}_b U'$, where $U$ is a rotation potential. This allows to simplify the quasilocal angular momentum term from \eqref{masang1}, i.e.
\begin{equation}
     \label{angularterm}
     \oint_{\boun} N^a A_a d S  = \oint_{\boun} U \epsilon^{ab} \slashed{\nabla}_a N_b d S .
\end{equation}
The spinorial counterpart of the volume element $\epsilon_{ab}$ of $\boun$ is
$\epsilon_{ABCD} = \frac{i}{2} \left(\epsilon_{AC} \gamma_{BD} + \epsilon_{BD} \gamma_{AC} \right)$,
and
\begin{equation}
\label{leviderN}
    \epsilon^{ab} \slashed{\nabla}_a N_b = \epsilon^{ABCD} \slashed{\nabla}_{AB}N_{CD}  =  \frac{2\lambda}{\Omega} \gamma^{AB}\phi_A \widehat{\phi}_B=\frac{2\lambda}{\Omega}(|\phi_0|^2-|\phi_1|^2),
\end{equation}
where the definition \eqref{spindivfreevector} has been taken into account. Inserting this expression into (\ref{angularterm}) and using $dS=\Omega^{2}d\mathbb{S}^2$ yields
\begin{equation}
\begin{split}
    \oint_{\boun} N^a A_a dS = 2 \lambda  \oint_{\mathbb{S}^2} U \left( |\widetilde{\phi}_0|^2 - |\widetilde{\phi}_1|^2 \right)   d\mathbb{S}^2,  \label{angularmomenterm}
\end{split}
\end{equation}
where $\widetilde{\phi}_{\bm A} =\sqrt{\Omega}\,\phi_{\bm A}$. The relation between the volume elements of $\boun$ and $\mathbb{S}^2$ can also be utilized to write the first term on the right-hand side of \eqref{masang1} in terms of an integral over $\mathbb{S}^2$. Ultimately,  

\begin{equation}
     \label{Sphericalmassinequality}
      4 \pi m \geq \sqrt{2} \oint \displaylimits_{\mathbb{S}^2} \rho '  |\widetilde{\phi}_0|^2\Omega d\mathbb{S}^2 + \frac{\kappa}{\sqrt{2}} O\left[\widetilde{\phi}, U \right],  
\end{equation}
where
$$
O \left[\widetilde{\phi}, U \right] \equiv - \frac{\lambda}{\kappa} \oint \displaylimits_{\mathbb{S}^2} U \left( |\widetilde{\phi}_0|^2 - |\widetilde{\phi}_1|^2 \right) d\mathbb{S}^2,
$$
i.e. the quasilocal angular momentum term can now be written only in terms of the geometry of $\mathbb{S}^2$ and the rotation potential $U$.
Note that the conformal factor $\Omega$ appears in the first term in the right-hand side of the above inequality and it is non-unique since the M\"obius group acting on $\mathbb{S}^2$ gives rise to different spherical metrics. The inequality obtained is therefore with respect to a given spherical metric, which can be thought of as a gauge choice here.

\section{Axisymmetric inner boundary and the Dirac-Killing system.}
\label{Secaxisym}

A natural assumption associated with the existence of a well-defined angular momentum is that the initial data is axisymmetric, i.e. there exists 1-form $v_a$ such that
$$
\nabla_{(a} v_{b)} = 0 \quad \mathrm{on} \quad \mathcal{S}.
$$
If the inner boundary $\boun$ is invariant under the action of the 1-parameter group of isometries generated by $v_a$, then $v_a = \Pi_a{}^bv_b$ ($v_a$ is intrinsic to $\boun$) and the projection of the Killing equation gives
$$
\slashed{\nabla}_{(a} v_{b)} =0 \implies \slashed{\nabla}_a v^a =0.
$$
This suggests that a natural choice for the vector $N^a$ defining the quasilocal angular momentum \eqref{szab} is that it arises as a solution to the boundary Killing equation, i.e. $\slashed{\nabla}_{(a} N_{b)}=0$. However, $N^a$ has already been constructed from a spinor $\phi_A$ satisfying a first-order Dirac-type equation \eqref{phigauge} on $\boun$. Hence, a natural question arises ---can such $N_a$ be also a Killing vector of the boundary? We will show that this cannot be the case on a generic $\boun$.

\medskip
In the sequel we will use an adapted system of coordinates $(\psi, \varphi)$ on the boundary, such that its metric $\sigma_{ab}$ can be written in the following form,
\begin{equation*}
    \sigma = - R^2\left(\tfrac{1}{F^2}d \psi\otimes d \psi +F^2 d \varphi\otimes d \varphi  \right), \quad \psi \in [\psi_0, \psi_1], \quad \varphi \in [0,2\pi],
\end{equation*}
where $F=F(\psi)$, $R$ is a constant and the axisymmetric Killing vector is now proportional to $\partial_{\varphi}$. To avoid the conical singularities on the poles we will assume that $F(\psi_0)=F(\psi_1)=0$. The NP operators $\delta$ and $\overline{\delta}$ reduce to
\begin{equation*}
    \delta=\tfrac{1}{\sqrt{2}R}\left( F\partial_{\psi}+\tfrac{i}{F}\partial_{\varphi}\right), \qquad \overline{\delta}=\tfrac{1}{\sqrt{2}R}\left( F\partial_{\psi}-\tfrac{i}{F}\partial_{\varphi}\right),
\end{equation*}
in this setting. Moreover, $\alpha-\overline{\beta}=\-\frac{1}{\sqrt{2}R}\partial_{\psi}F$ (see \cite{ColeRaczKroon2018} for details).

A straightforward computation yields
\begin{eqnarray*}
\slashed{\nabla}_{AB}N_{CD}+\slashed{\nabla}_{CD}N_{AB}&=&2\eth' \left( \overline{\phi}_0 \phi_1 \right)o_A o_B o_C o_D 
-\left( \eth \left( \overline{\phi}_0 \phi_1 \right)+\eth' \left(\phi_0 \overline{\phi}_1 \right)\right)o_A o_B \iota_C \iota_D \nonumber \\
&& -\left( \eth \left( \overline{\phi}_0 \phi_1 \right)+\eth' \left(\phi_0 \overline{\phi}_1 \right)\right)\iota_A \iota_B o_C o_D+2\eth \left( \phi_0 \overline{\phi}_1 \right)\iota_A \iota_B \iota_C \iota_D,  
\end{eqnarray*}
where we have used a decomposition of vector $N^a$ in terms of a spinor $\phi_A$ in accordance with (\ref{spindivfreevector}). Ultimately, the condition $\slashed{\nabla}_{(a} N_{b)} =0$ implies that
\begin{equation} \label{p0p1}
    \phi_0 \overline{\phi}_1 = i c F, \quad c\in\mathbb{R}.
\end{equation}

Additionally, the spinor $\phi_A$ satisfies a first-order Dirac-type equation \eqref{phigauge}, which can now be written as
\begin{equation} \label{direqc}
\begin{split}
    F \partial_{\psi} \phi_1 + \tfrac{\phi_1}{2} \partial_{\psi} F - \tfrac{i \sqrt{2} \lambda R}{\Omega} \phi_0 =0, \\
    F \partial_{\psi} \phi_0 + \tfrac{\phi_0}{2} \partial_{\psi} F - \tfrac{i \sqrt{2} \lambda R}{\Omega} \phi_1  =0.
\end{split}
\end{equation}
After multiplying the first equation by $\overline{\phi}_1$ and the second by $\overline{\phi}_0$ and performing some manipulations one arrives at 
\begin{equation*}
\begin{split}
     \partial_{\psi}  \left( F |\phi_1|^2 \right) + \tfrac{2 \sqrt{2} \lambda R c F}{\Omega} =0, \\
     \partial_{\psi}  \left( F |\phi_0|^2 \right) - \tfrac{2 \sqrt{2} \lambda R c F}{\Omega} =0,
\end{split}
\end{equation*}
where \eqref{p0p1} has been used. Hence
\begin{equation}
    \label{componentsnorm}
    |\phi_0|^2 = \frac{c_0}{F} + \frac{2 \sqrt{2} \lambda R c  }{F} \int \displaylimits_{\psi_0}^{\psi} \frac{F}{\Omega} dz, \quad |\phi_1|^2 = \frac{c_1}{F} - \frac{2 \sqrt{2} \lambda R c }{F} \int \displaylimits_{\psi_0}^{\psi} \frac{F}{\Omega} dz,
\end{equation}
for some $c_0, c_1\in \mathbb{R}$. On the other hand, one can apply $F \partial_{\psi}$ to both sides of \eqref{p0p1} and use \eqref{direqc} to get
\begin{equation*}
    \sqrt{2} \lambda R\left( |\phi_1|^2 - |\phi_0|^2 \right) =  2c \Omega F \partial_{\psi} F.
\end{equation*}
After using (\ref{componentsnorm}) we obtain the following compatibility condition for $F$,
\begin{equation}
    \label{Fcompatcondition}
    \sqrt{2} \lambda R \bigg( \frac{c_1-c_0}{F} - \frac{4 \sqrt{2} \lambda R c }{F} \int \displaylimits_{\psi_0}^{\psi} \frac{F}{\Omega} dz \bigg) = 2c \Omega F \partial_{\psi} F.
\end{equation}
Multiplying the above relation by $F$ and differentiating with respect to $\psi$ we arrive at
\begin{equation} 
\label{compF}
    c \left[ \partial_{\psi} \left( \Omega F^2 \partial_{\psi} F\right) + 4 R^2 \lambda^2 \Omega^{-1} F \right]=0,
\end{equation}
where $F(\psi_0)=0$ has been used. This equation implies that the metric of the inner boundary (via the function $F$) depends on the choice of the eigenvalue $\lambda$. This cannot be the case, as the former arises as part of the fixed geometric data associated with the initial hypersurface and the latter from the first-order Dirac-type equation, which is an auxiliary condition used to simplify the mass inequality. Hence, the only way to solve \eqref{compF} is to assume that $c=0$. In this case $\phi_A =0$ and the right-hand side of the mass-quasilocal angular momentum inequality \eqref{masang1} vanishes.

\begin{remark}
    In case of axisymmetric initial data, the inequality (\ref{Sphericalmassinequality}) reduces to the positivity of the ADM mass because $\phi_A$ vanishes. This suggests that the Szabados's quasi-local angular momentum cannot give rise to an ADM angular momentum appearing in the Penrose inequality.
\end{remark}

\section{Conclusions}
We have obtained a new bound for the ADM mass of \emph{asymptotically Schwarzschildean} initial data for the vacuum Einstein field equations in terms of the future inner null expansion of the inner MFTS boundary and its quasilocal angular momentum. Our approach bears similarities to the one presented in \cite{KopinskiKroon2021}, but we extend it here to allow for boundaries with nontrivial extrinsic geometry. An expression for quasilocal angular momentum (in the sense of Szabados \cite{Sza06}) has been recovered in the bound for the ADM mass by assuming a specific type of boundary condition for the approximate twistor equation-- a spinor $\phi_A$ solving a first-order Dirac-type equation. The strategy developed in this work could also be applied to obtain Penrose-type inequalities with different type of asymptotics (e.g. asymptotically hyperbolic) ---as long as the concept of quasilocal angular momentum is well-defined.

\medskip
\noindent
\textbf{Acknowledgments.} This project began during JK's visit to the School of Mathematical Sciences at Queen Mary University of London. He would like to express his gratitude to the school for its hospitality and acknowledge the support provided by the Polish National Science Centre through the MINIATURA project No. 2021/05/X/ST2/01151.


\begin{thebibliography}{}

\bibitem{AlaeeKunduri2023}
A.~Alaee and H.~Kunduri.
\newblock A Penrose-Type Inequality with Angular Momenta for Black Holes with 3-Sphere Horizon Topology
\newblock {\em J. Geom. Anal.}, 33, 231, 2023.


\bibitem{Anglada2018}
P.~Anglada.
\newblock Penrose-like inequality with angular momentum for minimal surfaces.
\newblock {\em Class. Quantum Grav.}, 35(4):045018, 2018.



\bibitem{Anglada2020}
P.~Anglada.
\newblock Comments on Penrose inequality with angular momentum for outermost apparent horizons.
\newblock {\em Class. Quantum Grav.}, 37.6:065023, 2020.



\bibitem{Ash91}
A.~Ashtekar.
\newblock {\em Lectures on non-perturbative canonical gravity}.
\newblock World Scientific, 1991.


\bibitem{BaeVal11a}
T.~B\"{a}ckdahl and J.~A. {Valiente Kroon}.
\newblock Approximate twistors and positive mass.
\newblock {\em Class. Quantum Grav.}, 28:075010, 2011.

\bibitem{Bartnik86}
R.~Bartnik.
\newblock The mass of an asymptotically 
flat manifold.
\newblock {\em Comm. PureAppl. Math.}, 39(5), 661-693, 1986.


\bibitem{ColeRaczKroon2018}
M.J. ~~Cole, I.~R\'acz, J.~A. Valiente Kroon
\newblock Killing spinor data on distorted black hole horizons and the uniqueness of stationary vacuum black holes.
\newblock {\em Class. Quantum Grav.}, 35(20), 2018.



\bibitem{Dai06}
S.~Dain.
\newblock Elliptic systems.
\newblock In J.~Frauendiener, D.~Giulini, and V.~Perlick, editors, {\em
  Analytical and Numerical approaches to General Relativity}, volume 692 of
  {\em Lect. Notes. Phys.}, page 117. Springer Verlag, 2006.




\bibitem{Dain2014}
S.~Dain.
\newblock Geometric inequalities for black holes.
\newblock {\em Gen. Rel. Grav.}, 46:1-23, 2014.




\bibitem{DainGabachC2018}
S.~Dain, M.~E. Gabach-Clement
\newblock Geometrical inequalities bounding angular momentum and charges in General Relativity.
\newblock {\em Living Reviews in Relativity}, 21(5), 2018.


\bibitem{GibHawHorPer83}
G.~W. Gibbons, S.~W. Hawking, G.~T. Horowitz, and M.~J. Perry.
\newblock Positivite mass theorems for black holes.
\newblock {\em Comm. Math. Phys.}, 88:295, 1983.


\bibitem{JaraczKhuri2018}
J.~Jaracz and M.~Khuri.
\newblock Bekenstein bounds, Penrose inequalities and black hole formation. 
\newblock {\em Physical Review D}, 97, 124026, 2018.


\bibitem{Jost2006}
J.~Jost.
\newblock {\em Compact Riemann surfaces}.
\newblock Springer-Verlag, 2006.


\bibitem{KopinskiKroon2021}
J.~Kopi\'nski, J.~A. Valiente Kroon
\newblock New spinorial approach to mass inequalities for black holes 
in general relativity.
\newblock {\em Phys. Rev. D}, 103, 2021.

\bibitem{KopinskiTafel2020}
J.~Kopi\'nski, J.~Tafel.
\newblock The Penrose inequality for perturbations of the Schwarzschild initial data.
\newblock {\em Class. Quantum Grav.}, 37 015012, 2020.

\bibitem{KopinskiTafel20202}
J.~Kopi\'nski, J.~Tafel.
\newblock The Penrose inequality for nonmaximal perturbations of the Schwarzschild initial data.
\newblock {\em Class. Quantum Grav.}, 37 105006, 2020.


\bibitem{KulczyckiMalec2021}
W.~Kulczycki, E.~Malec.
\newblock Extensions of the Penrose inequality with angular momentum.
\newblock {\em Physical Review D}, 103.6:064025, 2021


\bibitem{LudVick82}
M.~Ludvigsen, J.~A.~G. Vickers.
\newblock A simple proof of the positivity of the Bondi mass.
\newblock {\em J. Phys. A: Math. Gen.}, 15.2, L67-L70, 1982.


\bibitem{Mars2009}
M.~Mars.
\newblock Present status of the Penrose inequality 
\newblock {\em Class. Quantum Grav.}, 26, 193001, 2009.


\bibitem{Pen73}
R.~Penrose.
\newblock Naked singularities.
\newblock {\em Ann. N. Y. Acad. Sci.}, 224, 1973.



\bibitem{PenRin84}
R.~Penrose and W.~Rindler.
\newblock {\em Spinors and space-time. {V}olume 1. {T}wo-spinor calculus and
  relativistic fields}.
\newblock Cambridge University Press, 1984.




\bibitem{SchoenYau79}
R.M.~Schoen, S.T.~Yau.
\newblock Complete manifolds with nonnegative scalar curvature and the positive action conjecture in general relativity.
\newblock {\em Proc. Nat. Acad. Sci.}, 76(3), 1024-1025, 1979.


\bibitem{SchoenYau79bis}
R.M.~Schoen, S.T.~Yau.
\newblock On the proof of the positive mass conjecture
in general relativity.
\newblock {\em Comm. Math. Phys.}, 65, 45-76, 1979.



\bibitem{SchoenYau81}
R.M.~Schoen, S.T.~Yau.
\newblock The energy and the linear momentum of space-times in general relativity.
\newblock {\em Comm. Math. Phys.}, 79, 47-51, 1981.



\bibitem{Som80}
P.~Sommers.
\newblock Space spinors.
\newblock {\em J. Math. Phys.}, 21:2567, 1980.


\bibitem{Sza94a}
L.~B. Szabados.
\newblock Two-dimensional Sen connections in general relativity.
\newblock {\em Class. Quantum Grav.}, 11:1833, 1994.


\bibitem{Sza06}
L.~B. Szabados.
\newblock On a class of $2$-surface observables in general relativity.
\newblock {\em Class. Quantum Grav.}, 23, 2291-2302, 2006.




\bibitem{Sza08}
L.~B. Szabados.
\newblock Total angular momentum from dirac eigenspinors.
\newblock {\em Class. Quantum Grav.}, 25:025007, 2008.


\bibitem{CFEBook}
J.~A. {Valiente Kroon}.
\newblock {\em Conformal Methods in General Relativity}.
\newblock Cambridge University Press, 2016.


\bibitem{Wit81}
E.~Witten.
\newblock A new proof of the positive energy theorem.
\newblock {\em Comm. Math. Phys.}, 80:381, 1981.



\bibitem{WloRowLaw95}
J.~Wloka, B.~Rowley, and B.~Lawruk.
\newblock {\em Boundary value problems for elliptic systems}.
\newblock Cambridge University Press, 1995.








\end{thebibliography}
\end{document}